\documentclass{article}
\usepackage{amssymb}
\usepackage{amsfonts}
\usepackage{amsmath}

\input{tcilatex}

\begin{document}

\title{(1+1)-Dirac particle with position-dependent mass in complexified
Lorentz scalar interactions: effectively $\mathcal{PT}$-symmetric.}
\author{Omar Mustafa$^{1}$ and S.Habib Mazharimousavi$^{2}$ \\
Department of Physics, Eastern Mediterranean University, \\
G Magusa, North Cyprus, Mersin 10,Turkey\\
$^{1}$E-mail: omar.mustafa@emu.edu.tr\\
$^{2}$E-mail: habib.mazhari@emu.edu.tr}
\maketitle

\begin{abstract}
The effect of the built-in supersymmetric quantum mechanical language on the
spectrum of the (1+1)-Dirac equation, with position-dependent mass (PDM) and
complexified Lorentz scalar interactions, is re-emphasized. The signature of
the "quasi-parity" on the Dirac particles' spectra is also studied. A Dirac
particle with PDM and complexified scalar interactions of the form $S\left(
z\right) =S\left( x-ib\right) $ (an inversely linear plus linear, leading to
a $\mathcal{PT-}$symmetric oscillator model), and $S\left( x\right)
=S_{r}\left( x\right) +iS_{i}\left( x\right) $ (a $\mathcal{PT}$-symmetric
Scarf II model) are considered. Moreover, a first-order intertwining
differential operator and an $\eta $-weak-pseudo-Hermiticity generator are
presented and a complexified $\mathcal{PT}$-symmetric periodic-type model is
used as an illustrative example.

\end{abstract}

\section{Introduction}

A fermion bound to move in the $x$-direction (i.e., $p_{y}=p_{z}=0$)
mandates the decomposition of the (3+1)-dimensional Dirac equation into two
(1+1)-dimensional equations with two-component spinors and $2\times 2$ Pauli
matrices. Whilst the scalar, $S(x)$, and vector, $V(x)$, potentials preserve
their Lorentz structures (i.e., the former is added to the mass term of
Dirac equation while the minimal coupling is used, as usual, for the
latter), the angular momentum and spin are absent in the process.
Manifesting, in effect, a mathematically easily assessable and physically
more transparent exploration of the (1+1)-Dirac world.

Nevertheless, the supersymmetric quantum mechanical terminology is realized
(cf., e.g. [1-4]) as a \emph{hidden/built-in symmetry} in the
(1+1)-dimensional Dirac equation with "the mainly motivated by the MIT bag
model of quarks" Lorentz scalar potential [5]. For example, Nogami and
Toyama [1] have reported that the associated supersymmetric Schr\"{o}dinger
Hamiltonians $H_{1}$ and $H_{2}$ share the same energy spectrum including
the lowest states unless Dirac equation allows a zero-mode (i.e.,
zero-energy bound-state). Moreover, Jackiw and Rebbi [4] have reported that
if the Lorentz scalar potential is localized (i.e., $S(x)\rightarrow 0$ for $%
x\rightarrow \pm \infty $) no zero-mode is allowed. That is, only for some
Lorentz scalar potentials exhibiting certain \emph{topological }trends,
Dirac equation admits zero-mode.

Although the practical/experimental determination of the full spectrum is
often proved impossible, exact solvability of quantum mechanical models
(relativistic and non-relativistic) remains inviting and desirable. On the 
\emph{exact-solvability} methodical side, however, attention was (by large)
paid to the non-relativistic Schr\"{o}dinger equation, whereas the
relativistic Klein-Gordon and Dirac equations are left unfortunates. Not
only within the recent revival of the unusual non-Hermitian complexified
Hamiltonians' settings [6-13] , but also within the usual Hermitian ones
including those with position-dependent mass (PDM) [14-19].

In their pioneering generalization of the non-relativistic quantization
recipe (i.e., $\mathcal{PT}$-symmetric Hamiltonians, where $\mathcal{P}$
denotes parity and the complex conjugation $\mathcal{T}$ \ mimics time
reversal.), Bender and Boettcher [6] have suggested a tentative
weakening/relaxation of Hermiticity as a necessary condition for the reality
of the spectrum (i.e., the reality of the spectrum is secured by the
exactness of $\mathcal{PT}$ -symmetry). However, Mostafazadeh [8] has
introduced a broader class of the so-called pseudo-Hermitian Hamiltonians
with real spectra (within which $\mathcal{PT}$ -symmetric Hamiltonians form
a subclass). He has, basically, advocated the "user-friendly" consensus that
neither Hermiticity nor $\mathcal{PT}$ -symmetry serve as necessary
conditions for the reality of the spectrum of a quantum Hamiltonian [6-13].
Yet, the existence of the real eigenvalues is realized to be associated\
with a non-Hermitian Hamiltonian provided that it is an $\eta $%
-pseudo-Hermitian, $\eta \,H=H^{\dagger }\,\eta $, with respect to the
nontrivial "metric" intertwining operator $\eta $ ($=O^{\dagger }O$, for
some linear invertible operator $O:\mathcal{H}{\small \rightarrow }\mathcal{H%
}$, where $\mathcal{H}$ is the Hilbert space). Furthermore, one may rather
choose to be disloyal to the Hermiticity (cf., e.g., Bagchi and Quesne
[10]), and "linear" and/or "invertible" (cf., e.g., Solombrino [11] and
Mustafa and Mazharimousavi [12]) conditions on the intertwiner $\eta $, and
hence relaxing $H$ to be an $\eta $-weak-pseudo-Hermitian.

On the one (among others, some of which are readily mentioned above) of the
main stimulants/inspirations of the present article, we may re-collect that
quantum particles endowed with PDM constitute useful models for the study of
many physical problems. In particular (but not limited to), they are used in
the energy density many-body problem, in the determination of the electronic
properties of semiconductors and quantum dots [cf., e.g., the sample of
references in [12-19]), etc.

In the forthcoming text, we shall focus (in addition to the (1+1)-Dirac
particle with PDM in complexified Lorentz scalar interactions) on two main
spectral phenomenological properties. Namely the energy-levels crossings
(manifested by the "quasi-parity" settings of Znojil's [10]
attractive/repulsive-like core) and the related effects to the
hidden/built-in supersymmetric terminology in the (1+1)-Dirac equation. Both
in the usual Hermitian and the unusual complexified non-Hermitian settings.

The organization of this article is in order. In section 2, we discuss the
(1+1)-Dirac equation with PDM and a Lorentz-scalar interaction and
re-emphasize Nogami's and Toyama's [1] hidden/built-in supersymmetric
language. We report, in section 3, some consequences of a complexified
non-Hermitian $\mathcal{PT}$ -symmetric Lorentz scalar potentials belonging
to two different classes: $S\left( x\right) \rightarrow S\left( x-ib\right)
=S\left( z\right) ;$ $x,b\in 
\mathbb{R}
,$ $z\in 
\mathbb{C}
$ and $S\left( x\right) =S_{r}\left( x\right) +iS_{i}\left( x\right)
;S_{r}\left( x\right) ,S_{i}\left( x\right) \in 
\mathbb{R}
$; an inversely linear plus linear and a Scarf II models, respectively. In
section 4, we explore one possibility of $\eta $-weak-pseudo-Hermiticity
generators via a first-order intertwining differential operator. We
exemplify this possibility by an $\eta $-weak-pseudo-Hermitian $\mathcal{PT}$
-symmetric periodic-type model. We conclude in section 5.

\section{(1+1)-Dirac equation with a position dependent mass and a Lorentz
Scalar interactions}

In the presence of a time-independent position-dependent mass , $m\left(
x\right) $, and a Lorentz scalar interaction, $S\left( x\right) $, the
(1+1)-dimensional time-independent Dirac equation (in $c=\hbar =1$ units)
reads 
\begin{equation}
H_{D}\Psi \left( x\right) =E\Psi \left( x\right) \text{ };\text{ \ \ }%
H_{D}=\alpha p+\beta \left[ m\left( x\right) +S\left( x\right) \right] ,
\end{equation}%
where $p=-i\partial _{x},$ $\alpha $ and $\beta $ are the usual $2\times 2$
Pauli matrices satisfying the relations $\alpha ^{2}=\beta ^{2}=1$ and $%
\left\{ \alpha ,\beta \right\} =0$, $E$ is the energy of the Dirac particle,
and $\Psi \left( x\right) $ is the two-component spinor. Equation (1) with 
\begin{equation}
\Psi \left( x\right) =\left( 
\begin{array}{c}
\psi _{+}\left( x\right) \\ 
\psi _{-}\left( x\right)%
\end{array}%
\right) \text{ },\text{ }\alpha =\sigma _{2}=\left( 
\begin{array}{cc}
0 & -i \\ 
i & 0%
\end{array}%
\right) \text{ },\text{ }\beta =\sigma _{1}=\left( 
\begin{array}{cc}
0 & 1 \\ 
1 & 0%
\end{array}%
\right)
\end{equation}%
reduces (with $M\left( x\right) =m\left( x\right) +S\left( x\right) $) to%
\begin{equation}
\left( 
\begin{array}{cc}
0 & -\partial _{x}+M\left( x\right) \\ 
\partial _{x}+M\left( x\right) & 0%
\end{array}%
\right) \left( 
\begin{array}{c}
\psi _{+}\left( x\right) \\ 
\psi _{-}\left( x\right)%
\end{array}%
\right) =E\left( 
\begin{array}{c}
\psi _{+}\left( x\right) \\ 
\psi _{-}\left( x\right)%
\end{array}%
\right)
\end{equation}%
which decouples, in turn, into%
\begin{gather}
\left[ -\partial _{x}+M\left( x\right) \right] \psi _{-}\left( x\right)
=E\psi _{+}\left( x\right) ,\medskip \\
\left[ \partial _{x}+M\left( x\right) \right] \psi _{+}\left( x\right)
=E\psi _{-}\left( x\right) .\medskip
\end{gather}%
This would, with $\omega =\pm 1$, imply a Schr\"{o}dinger-like equation%
\begin{gather}
\left\{ -\partial _{x}^{2}+V_{\omega }\left( x\right) \right\} \psi _{\omega
}\left( x\right) =\lambda _{\omega }\psi _{\omega }\left( x\right) ;\medskip
\\
V_{\omega }\left( x\right) =M\left( x\right) ^{2}-\omega M^{\prime }\left(
x\right) \text{ };\text{ }\lambda _{\omega }=E_{\omega }^{2},\medskip
\end{gather}%
where prime denotes derivative with respect to $x$ (i.e., $\partial _{x}$).
Nevertheless, a built-in supersymmetric quantum mechanical language is
obvious in equation (7). That is, if the superpotential is defined as $%
W\left( x\right) =-M\left( x\right) $, \ then the supersymmetric partner
potentials are given by%
\begin{equation*}
V_{\omega }\left( x\right) =M\left( x\right) ^{2}-\omega M^{\prime }\left(
x\right) =W^{2}\left( x\right) \pm W^{\prime }\left( x\right) .
\end{equation*}%
In this case, one would label $H_{-}=$ $-\partial _{x}^{2}+V_{-}\left(
x\right) $ and $H_{+}=$ $-\partial _{x}^{2}+V_{+}\left( x\right) $ as the
two partner Hamiltonians (cf., e.g., Alhaidari [16], and Sinha and Roy [3]).
Of course, such supersymmetric language would leave its
fingerprints/signature on the spectrum, as shall be witnessed in the
forthcoming experiments with both Hermitian and non-Hermitian models.

\section{Consequences of complexified non-Hermitian Lorentz scalar
interactions}

In this section, we consider two cases: a Dirac particle with $S\left(
x\right) \rightarrow S\left( x-ib\right) =S\left( z\right) $ (where $%
\mathbb{C}
\ni z=x-ib;$ $%
\mathbb{R}
\ni x\in \left( -\infty ,\infty \right) $, and $%
\mathbb{R}
\ni \func{Im}z=-b<0$, i.e., a simple constant downward shift of the
coordinate is considered), and Dirac particle with $S\left( x\right)
=S_{r}\left( x\right) +iS_{i}\left( x\right) $, where $S_{r}\left( x\right)
,S_{i}\left( x\right) \in 
\mathbb{R}
$.

\subsection{Dirac particle with a complexified Lorentz scalar $S\left(
x\right) \rightarrow S\left( z\right) \equiv S\left( x-ib\right) $: a $%
\mathcal{PT-}$symmetrized Znojil's oscillator}

A Dirac particle under the influence of $M\left( z\left( x\right) \right)
=m\left( x\right) +S\left( z\left( x\right) \right) $ (where $z=x-ib;$ $x\in
\left( -\infty ,\infty \right) $, and $%
\mathbb{R}
\ni \func{Im}x=-b<0$, i.e., a simple constant downward shift of the
coordinate is considered) would result in recasting (6) and (7) as%
\begin{equation}
\left\{ -\partial _{z}^{2}+V_{\omega }\left( z\right) \right\} \psi _{\omega
}\left( z\right) =\lambda _{\omega }\psi _{\omega }\left( z\right) ;\text{ }%
V_{\omega }\left( z\right) =M\left( z\right) ^{2}-\omega M^{\prime }\left(
z\right) .
\end{equation}

Then, a Dirac particle endowed with a mass function of the form $m(x)=Bx/4;$ 
$%
\mathbb{R}
\ni B\geq 0,$ under the influence of a complexified non-Hermitian Lorentz
scalar interaction $S\left( z\right) =\frac{B}{4}z+\frac{A}{z}-i\frac{B}{4}b$
would imply 
\begin{equation}
M\left( z\right) =\frac{B}{2}z+\frac{A}{z}.
\end{equation}%
Which, in effect, yields two complexified non-Hermitian $\mathcal{PT-}$%
symmetric partner potentials%
\begin{equation}
V_{\omega }\left( z\right) =\frac{1}{4}B^{2}z^{2}+\frac{A\left( A+\omega
\right) }{z^{2}}+B\left( A-\frac{\omega }{2}\right) \medskip .
\end{equation}%
Moreover, it should be noted that $V_{+}\left( z\right) $ represents a $%
\mathcal{PT-}$symmetric complexified oscillator perturbed by a "shifted by a
constant" Znojil's [10] repulsive/attractive core, i. e., with the
parametric choice $A=\alpha -\frac{1}{2};$ $\alpha \geq 0$, one gets%
\begin{equation}
\frac{A\left( A+1\right) }{z^{2}}+B\left( A+\frac{1}{2}\right) =\frac{\left(
\alpha ^{2}-\frac{1}{4}\right) }{z^{2}}+B\left( A+\frac{1}{2}\right) ,
\end{equation}%
cf., e.g. Mustafa and Znojil [7], and $V_{-}\left( z\right) $ represents a $%
\mathcal{PT-}$symmetric oscillator perturbed by a shifted/rescaled, say,
Znojil's [10] repulsive/attractive core, i.e.,%
\begin{equation}
\frac{A\left( A-1\right) }{z^{2}}+B\left( A-\frac{1}{2}\right) =\frac{\left(
\alpha ^{2}-\frac{1}{4}\right) -2\left( \alpha -\frac{1}{2}\right) }{z^{2}}%
-\alpha B.
\end{equation}%
Under these settings, one would map Znojil's results [10], taking into
account our discussion on the supersymmetric-like partner potentials in
(10), and obtain%
\begin{eqnarray}
\lambda _{+,q} &=&\frac{B}{2}\left( 4n+2q\alpha +2\alpha \right) \medskip
\medskip ;\text{ }n=0,1,2,\cdots ,\medskip \\
\lambda _{-,q} &=&\frac{B}{2}\left[ 4n+\left( \alpha +1\right) \left(
2-2q\right) \right] ;\text{ }n=0,1,2,\cdots .\medskip
\end{eqnarray}%
We observe the supersymmetric language \emph{"signature"} \ in $\lambda
_{+,q=+1}=\lambda _{-,q=+1}$ for even quasi-parity and $\lambda
_{+,q=-1}+const.=\lambda _{-,q=-1}$ for odd quasi-parity, i.e., 
\begin{equation}
\lambda _{+,q=+1}=\lambda _{-,q=+1}=2B\left( n+\alpha \right) ,
\end{equation}%
and%
\begin{equation}
\lambda _{-,q=-1}=\lambda _{+,q=-1}+2B=2B\left( n+1\right) .\medskip
\end{equation}%
Leading, in effect, (with $E_{+,q}=+\sqrt{\lambda _{+,q}}$ and $E_{-,q}=-%
\sqrt{\lambda _{-,q}}$) to 
\begin{eqnarray}
E_{+,q} &=&\left\{ 
\begin{tabular}{l}
$E_{+,q=+1}=+\sqrt{2B\left( n+\alpha \right) }\medskip $ \\ 
$E_{+,q=-1}=+\sqrt{2Bn}\medskip $%
\end{tabular}%
\right. , \\
E_{-,q} &=&\left\{ 
\begin{tabular}{l}
$E_{-,q=+1}=-\sqrt{2B\left( n+\alpha \right) }\medskip $ \\ 
$E_{-,q=-1}=-\sqrt{2B\left( n+1\right) }\medskip $%
\end{tabular}%
\right. .
\end{eqnarray}%
Yet, the \emph{energy-levels crossing} phenomenon (a quasi-parity signature
on the spectrum above) is also observed unavoidable. That is, the two sets
of energies in (17) cross with each other when%
\begin{equation}
E_{+}\left( n=n_{1},q=+1\right) =E_{+}\left( n=n_{2},q=-1\right)
\Longrightarrow n_{2}-n_{1}=\alpha
\end{equation}%
and the sets of energies in (18) cross with each other when%
\begin{equation}
E_{-}\left( n=n_{3},q=+1\right) =E_{-}\left( n=n_{4},q=-1\right)
\Longrightarrow n_{4}-n_{3}=\alpha -1.
\end{equation}

\subsection{Dirac particle with a complexified Lorentz scalar $S\left(
x\right) =S_{r}\left( x\right) +iS_{i}\left( x\right) $: a $\mathcal{PT}$%
-symmetric Scarf II model}

In this section we consider a class of a complexified Lorentz scalar models
of the form $S\left( x\right) =S_{r}\left( x\right) +iS_{i}\left( x\right) $
and position-dependent mass $m\left( x\right) \neq 0$ in $M\left( x\right)
=m\left( x\right) +S\left( x\right) $. For simplicity of calculations, we
take%
\begin{equation}
M\left( x\right) =\tilde{M}\left( x\right) +iS_{i}\left( x\right) \text{ };%
\text{ }\tilde{M}\left( x\right) =m\left( x\right) +S_{r}\left( x\right)
\end{equation}

If we assume that $\tilde{M}\left( x\right) =\left( A+B\right) \tanh x$ and $%
S_{i}\left( x\right) =-\left( A-B\right) \func{sech}x$ then%
\begin{equation}
M\left( x\right) =\left( A+B\right) \tanh x-i\left( A-B\right) \func{sech}x
\end{equation}%
and consequently the corresponding supersymmetric $\mathcal{PT}$-symmetric
partner potentials are given by%
\begin{equation}
V_{\pm }\left( x\right) =-C_{1}\func{sech}^{2}x-iC_{2}\func{sech}x\tanh
x+\left( A+B\right) ^{2}
\end{equation}%
where $C_{1}=2\left( A^{2}+B^{2}\right) +\omega \left( A+B\right) $ and $%
C_{2}=\left( 2A+2B+\omega \right) \left( A-B\right) $. It is obvious that $%
V_{\pm }\left( x\right) $ is the well known complexified $\mathcal{PT}$%
-symmetric Scarf II model. Moreover, it should be noted that $V_{+}\left(
x\right) $ and $V_{-}\left( x\right) $ imitate the pseudo-supersymmetric $%
\mathcal{PT}$-symmetric partner potentials $U_{2}\left( x\right) $ and $%
U_{1}\left( x\right) $, respectively, reported in Eq.s (38)-(40) by Sinha
and Roy [3]. The solution of which can be easily mapped into the above
model, by taking the constant mass in Sinha and Roy [3] equals zero, to
obtain%
\begin{equation}
E_{+,n}\left( A,B\right) =+\sqrt{2\left( A+B\right) \left( n+1\right)
-\left( n+1\right) ^{2}}\text{ };\text{ }n=0,1,2,\cdots .
\end{equation}%
\begin{equation}
E_{-,n}\left( A,B\right) =-\sqrt{2\left( A+B\right) n-n^{2}}\text{ };\text{ }%
n=0,1,2,\cdots .
\end{equation}%
However, it is obvious that%
\begin{equation}
E_{+,n}\left( A,B\right) \in 
\mathbb{R}
\iff \left[ 2\left( A+B\right) -1\right] \geq n
\end{equation}%
and%
\begin{equation}
E_{-,n}\left( A,B\right) \in 
\mathbb{R}
\iff 2\left( A+B\right) \geq n\ .
\end{equation}%
This result in effect documents the fact that $\mathcal{PT}$-symmetry is not
an enough condition to guarantee the reality of Dirac spectrum but rather it
should be complemented by the condition $E_{n}^{2}\geq 0$. Moreover, \emph{%
energy-levels crossing} phenomenon introduces itself (in this case, of
course, not as a quasi-parity effect but rather as a spectral property) in
the following scenario: the energy levels in the set (24) perform \emph{%
energy-levels crossing} among each other when%
\begin{equation}
E_{+,n_{1}}\left( A,B\right) =E_{+,n_{2}}\left( A,B\right) \Longrightarrow
n_{1}+n_{2}=2\left( A+B-1\right) ,
\end{equation}%
and similar trend is also obvious in (25) when%
\begin{equation}
E_{-,n_{3}}\left( A,B\right) =E_{-,n_{4}}\left( A,B\right) \Longrightarrow
n_{3}+n_{4}=2\left( A+B\right) .
\end{equation}

\section{Consequences of $\protect\eta $-weak-pseudo-Hermiticity via a
first-order intertwiner}

A complexified non-Hermitian Lorentz scalar interaction, $S\left( x\right)
=S_{r}\left( x\right) +iS_{i}\left( x\right) $, where $S_{r}\left( x\right)
,S_{i}\left( x\right) \in 
\mathbb{R}
$, would result in%
\begin{equation}
\func{Re}V_{\pm }\left( x\right) =m\left( x\right) ^{2}+S_{r}\left( x\right)
^{2}-S_{i}\left( x\right) ^{2}+2m\left( x\right) S_{r}\left( x\right)
-\omega \left[ m^{\prime }\left( x\right) +S_{r}^{\prime }\left( x\right) %
\right] ,
\end{equation}%
\begin{equation}
\func{Im}V_{\pm }\left( x\right) =2m\left( x\right) S_{i}\left( x\right)
+2S_{i}\left( x\right) S_{r}\left( x\right) -\omega S_{i}^{\prime }\left(
x\right)
\end{equation}%
We may now work with a Schr\"{o}dinger-like non-Hermitian Hamiltonian
operator $\tilde{H}_{\pm }=-\partial _{x}^{2}+V_{\pm }\left( x\right) $ with
the eigenvalues $\lambda _{\pm }=E^{2}$. Then $\tilde{H}_{\pm }$ is an $\eta 
$-weak-pseudo-Hermitian (admitting real eigenvalues $\lambda _{\pm
}=E^{2}\in 
\mathbb{R}
$) with respect to the first-order Hermitian intertwiner%
\begin{equation}
\eta =-i\,\,\partial _{x}+G\left( x\right) ,
\end{equation}%
where $G\left( x\right) \in 
\mathbb{R}
$, if it satisfies the intertwining relation $\eta \tilde{H}_{\pm }=\tilde{H}%
_{\pm }^{\dagger }\,\eta $ (it is not difficult to show that $\left( \eta 
\tilde{H}_{\pm }\right) $ is Hermitian too).

Under such $\eta $-weak-pseudo-Hermiticity settings, the intertwining
relation would result in 
\begin{equation}
\func{Im}V_{\pm }\left( x\right) =-G^{\prime }\left( x\right) ,\text{\ and \ 
}\func{Re}V_{\pm }\left( x\right) =-G\left( x\right) ^{2}
\end{equation}%
to yield, respectively,%
\begin{equation}
-G^{\prime }\left( x\right) =2m\left( x\right) S_{i}\left( x\right)
+2S_{i}\left( x\right) S_{r}\left( x\right) -\omega S_{i}^{\prime }\left(
x\right) ,
\end{equation}%
\begin{equation}
-G\left( x\right) ^{2}=m\left( x\right) ^{2}+S_{r}\left( x\right)
^{2}-S_{i}\left( x\right) ^{2}+2m\left( x\right) S_{r}\left( x\right)
-\omega \left[ m^{\prime }\left( x\right) +S_{r}^{\prime }\left( x\right) %
\right] .
\end{equation}%
Consequently, Eq.(34) implies%
\begin{equation}
m\left( x\right) +S_{r}\left( x\right) =\frac{-G^{\prime }\left( x\right)
+\omega S_{i}^{\prime }\left( x\right) }{2S_{i}\left( x\right) }.
\end{equation}%
Substituting (36) in (35) would yield%
\begin{equation}
\left[ \frac{-G^{\prime }\left( x\right) +\omega S_{i}^{\prime }\left(
x\right) }{2S_{i}\left( x\right) }\right] ^{2}-\omega \left[ \frac{%
-G^{\prime }\left( x\right) +\omega S_{i}^{\prime }\left( x\right) }{%
2S_{i}\left( x\right) }\right] ^{\prime }=\left[ \omega S_{i}\left( x\right) %
\right] ^{2}-G\left( x\right) ^{2}.
\end{equation}%
The simplest solution of which is given by (with $\omega =\pm 1$) the choice%
\begin{equation}
G_{\pm }\left( x\right) =\omega S_{i}\left( x\right) \implies S_{i,\pm
}\left( x\right) =\omega G\left( x\right) \implies S_{r}\left( x\right)
=-m\left( x\right) .
\end{equation}

In the forthcoming experiment, we shall be interested in the family of
complexified Lorentz scalar interactions of the form $S\left( x\right)
=-m\left( x\right) +iS_{i}\left( x\right) $. With such settings in point,
the Dirac Hamiltonian in (1) collapses into%
\begin{equation}
H_{D}=\sigma _{2}p+i\sigma _{1}S_{i}\left( x\right) =\left( 
\begin{array}{cc}
0 & -\partial _{x}+iS_{i}\left( x\right) \\ 
\partial _{x}+iS_{i}\left( x\right) & 0%
\end{array}%
\right) .
\end{equation}%
Consequently and without any loss of generality, one may very well recast
our $\eta $-weak-pseudo-Hermitian Schr\"{o}dinger-like Hamiltonian as%
\begin{equation}
\tilde{H}_{\pm }=-\partial _{x}^{2}+V_{\pm }\left( x\right) =-\partial
_{x}^{2}-G\left( x\right) ^{2}-i\omega G^{\prime }\left( x\right) .
\end{equation}

\subsection{An $\protect\eta $-weak-pseudo-Hermitian $\mathcal{PT}$%
-symmetric periodic-type model}

An $\eta $-weak-pseudo-Hermiticity generator of a periodic nature of the form

\begin{equation}
G(x)=-\frac{4}{3\cos ^{2}x-4}-\frac{5}{4}
\end{equation}%
would imply $\mathcal{PT}$-symmetric periodic-type effective potentials%
\begin{equation}
V_{\pm }\left( x\right) =-G(x)^{2}-i\omega G(x)^{\prime }=\frac{1}{9}\frac{%
-30\cos ^{2}x+24}{(\cos ^{2}x-\frac{4}{3})^{2}}+i\frac{4\omega \sin 2x}{%
3(\cos ^{2}x-\frac{4}{3})^{2}}-\frac{25}{16}
\end{equation}%
which, in a straightforward manner, can be rewritten as%
\begin{equation}
V_{\pm }\left( x\right) =-\frac{6}{(\cos x+2i\omega \sin x)^{2}}-\frac{25}{16%
}.
\end{equation}%
It should be noted here that $V_{+}\left( x\right) $ is the $\mathcal{PT}$%
-symmetric periodic-type effective potential representing a \emph{"shifted
by a constant"} \ Samsonov-Roy's [20] periodic potential model satisfying $%
V_{\pm }\left( x\right) =V_{\mp }\left( -x\right) $. Hence, if we defined 
\begin{equation}
\tilde{V}_{\pm }\left( x\right) =-\frac{6}{(\cos x+2i\omega \sin x)^{2}},
\end{equation}%
then (with $\mathcal{P}$ denoting parity)%
\begin{equation*}
\mathcal{P}\tilde{V}_{\pm }\left( x\right) =\tilde{V}_{\pm }\left( -x\right)
=-\tilde{V}_{\pm }\left( x\right) =\tilde{V}_{\mp }\left( x\right) .
\end{equation*}%
Consequently, $\tilde{V}_{\pm }\left( x\right) $ and $\tilde{V}_{\mp }\left(
x\right) $ mirror reflect each other. A result that provides a safe passage
through the transformation $x\longrightarrow y=-x$ and mandates%
\begin{equation*}
\tilde{H}_{\pm }=-\partial _{x}^{2}+V_{\pm }\left( x\right) =-\partial
_{y}^{2}+V_{\mp }\left( y\right)
\end{equation*}

The solution of which is reported for the interval $x\in \left( -\pi ,\pi
\right) $ (equivalently, $y\in \left( -\pi ,\pi \right) $) with the boundary
conditions $\psi _{n,\pm }\left( -\pi \right) =\psi _{n,\pm }\left( \pi
\right) =0$ as%
\begin{eqnarray}
\psi _{n,\pm }\left( x\right) &=&\left\{ \left[ \left( 16-n^{2}\right) \cos
x\mp 2i\left( n^{2}-4\right) \sin x\right] \sin \left[ \frac{n}{2}\left( \pi
\pm x\right) \right] \right. \medskip  \notag \\
&&\left. \mp 6n\sin x\cos \left[ \frac{n}{2}\left( \pi \pm x\right) \right]
\right\} (\cos x\pm 2i\sin x)^{-1}\medskip
\end{eqnarray}%
and%
\begin{equation}
E_{n,\pm }=\pm \sqrt{\lambda _{\pm }}=\pm \sqrt{\frac{n^{2}}{4}-\frac{25}{16}%
}\text{ };\text{ \ }n=3,4,5,\cdots .\medskip
\end{equation}%
It should be reported here that the values of $n<3$ are scarified for the
sake of the reality of the spectrum.

\section{Conclusion}

In this work, the effect of the built-in supersymmetric quantum mechanical
language on the structure of the decomposed (1+1)-Dirac equation, with PDM
and complexified Lorentz scalar interactions, is re-emphasized. In the
process, the signature of the "quasi-parity" (manifested by Znojil's
attractive/repulsive-like core [10]) is also studied. In so doing, a
"quasi-free" Dirac particle with PDM (an inversely linear plus linear), a
Dirac particle with PDM and complexified scalar interactions, $S\left(
z\right) =S\left( x-ib\right) ;$ $x,b\in 
\mathbb{R}
,$ $z\in 
\mathbb{C}
$ (an inversely linear plus linear, leading to a $\mathcal{PT-}$symmetric
oscillator model), and $S\left( x\right) =S_{r}\left( x\right) +iS_{i}\left(
x\right) ;S_{r}\left( x\right) ,S_{i}\left( x\right) \in 
\mathbb{R}
$ (a $\mathcal{PT}$-symmetric Scarf II model) are considered. Moreover, a
first-order intertwining differential operator and an $\eta $%
-weak-pseudo-Hermiticity generator are presented (a complexified $\mathcal{PT%
}$-symmetric periodic-type model is used).

In the light of our experience above we have observed that the associated
supersymmetric signature on the spectrum of the (1+1)-Dirac particle results
in exact-isospectral (i.e., including the lowest states) partner
Hamiltonians $H_{1}$ and $H_{2}$ for "even" quasi-parity, however, they
share the same energy spectrum with a "missing" lowest state for "odd"
quasi-parity. Nevertheless, we may report that the \emph{energy-levels
crossing} is only feasible among positive-energy states (i.e., above $E=0$)
or among negative-energy states (i.e., below $E=0$), at least as long as our
illustrative examples are concerned. We may also add that neither the
exactness of $\mathcal{PT}$-symmetry nor pseudo-Hermiticity are enough
conditions for the reality of the Dirac spectrum, they should be rather
complemented by the condition $%
\mathbb{R}
\ni E^{2}>0$. Finally, one may need to sacrifice some energy states for the
sake of the reality of the Dirac particle spectrum.

\vspace{0pt}

\end{document}